\documentclass[pre,tighten]{revtex4}

\usepackage{amssymb,amsmath}

\usepackage{graphicx}
\newcommand{\dd}{\text{d}}

\begin{document}

\title{Grad's Moment Method for a Low-Density Granular Gas. Navier-Stokes Transport Coefficients}



\author{Vicente Garz\'{o}}
\email{vicenteg@unex.es}
\homepage{URL: http://www.unex.es/eweb/fisteor/vicente/}
\affiliation{Departamento de F\'{\i}sica, Universidad de Extremadura, E-06071 Badajoz, Spain}

\begin{abstract}
The Navier-Stokes transport coefficients for a granular gas of smooth inelastic hard disks or spheres are determined from the inelastic Boltzmann equation by means of Grad's moment method. The shear viscosity $\eta$, the thermal conductivity $\kappa$ and the new transport coefficient $\mu$ (not present for elastic collisions) are explicitly obtained as nonlinear functions of the (constant) coefficient of restitution $\alpha$. The expressions of $\eta$, $\kappa$, and $\mu$ agree with those previously obtained from the Chapman-Enskog method by using the first Sonine approximation. A comparison with previous results  derived from Grad's moment method for two and three dimensions is also carried out.

\end{abstract}

\maketitle


\section{Introduction}
\label{sec1}

The simplest model for a granular gas is a system constituted by smooth hard spheres or disks with inelastic collisions. The loss of energy in each binary collision is accounted for a constant coefficient of normal restitution $\alpha \leq 1$. The essential difference from that for ordinary gases ($\alpha=1$) is the absence of energy conservation, yielding both obvious and subtle modifications of the usual Navier-Stokes (NS) equations for states with small spatial gradients of the hydrodynamic fields. For a low-density gas, the Boltzmann equation (BE) conveniently adapted to dissipative dynamics has been used in recent years as the starting point to derive the NS hydrodynamic equations. Therefore, assuming the existence of a \emph{normal} solution for sufficiently long space and time scales, the Chapman-Enskog method \cite{CC70} has been applied to determine the explicit forms of the relevant transport coefficients for dilute \cite{BDKS98} and moderately dense \cite{GD99} gases. As in the case of elastic collisions \cite{CC70}, the NS transport coefficients are given in terms of the solutions of a set of coupled linear integral equations which can be approximately solved by considering the lowest terms in a Sonine polynomial expansion. In this context, granular hydrodynamics follows similar steps as those made for ordinary gases.

An alternative way of solving the BE is by means of Grad's moment method \cite{G49}. The idea behind Grad's method is to expand the velocity distribution function $f$ in a complete set of orthogonal polynomials, the coefficients being the corresponding velocity moments of $f$. Then, the expansion is truncated at a given order $k$ so that, the exact distribution function $f$ is replaced by its truncated expansion $f^{(k)}$. When the approximation $f^{(k)}$  is substituted into the hierarchy of the moment equations up to order $k$, one obtains a closed set of coupled equations for the retained velocity moments.

The inelastic Enskog equation was solved years ago by Jenkins and Richman \cite{JR85a,JR85b} by applying Grad's thirteen-moment for a dense gas of inelastic hard spheres. Although the application of Grad's method to the Enskog or Boltzmann equations is not restricted to nearly elastic particles ($\alpha \simeq 1$), the results derived by Jenkins and Richman  \cite{JR85a,JR85b} neglect the cooling effects on the granular temperature $T$ due to the cooling rate. Given that this assumption can only be justified for nearly elastic systems, their expressions for the NS transport coefficients differ from those obtained in Ref.\ \cite{BDKS98} from the Chapman-Enskog method for \emph{arbitrary} degree of inelasticity.

The goal of this contribution is to apply Grad's method to get the NS transport coefficients of $d$-dimensional granular gases at low-density. In contrast to the Jenkins-
Richman theory \cite{JR85a,JR85b}, my calculations account for the time dependence of the temperature coming from the inelastic cooling. As a consequence, the results are not limited \emph{a priori} to weak inelastic systems. In particular, when only linear terms in the spatial gradients are retained (NS approximation), the constitutive relations for the pressure tensor $P_{ij}$ and the heat flux ${\bf q}$ are
\begin{equation}
\label{1}
P_{ij}=p\delta_{ij}-\eta \left(\partial_{i}U_{j}+
\partial_{j}U_{i}-\frac{2}{d}\delta_{ij}\nabla \cdot \mathbf{U}\right),
\end{equation}
\begin{equation}
\label{2}
\mathbf{q}=-\kappa \nabla T-\mu \nabla n,
\end{equation}
where $p=nT$ is the hydrostatic pressure, $\textbf{U}$ is the mean flow velocity, $T$ is the granular temperature, and $n$ is the number density. In addition, $\eta$ is the shear viscosity coefficient, $\kappa$ is the thermal conductivity coefficient and $\mu$ is a new transport coefficient not present for elastic collisions. In contrast to the Jenkins-Richman theory \cite{JR85a,JR85b}, the present results show that the expressions of $\eta$, $\kappa$ and $\mu$ obtained from Grad's moment method are the same as those obtained \cite{BDKS98} from the Chapman-Enskog expansion in the first Sonine approximation.

\section{Boltzmann kinetic theory for granular gases}
\label{sec2}

We consider a granular gas composed by hard disks ($d=2$) or spheres ($d=3$) of mass $m$ and diameter $\sigma$. In the simplest model, the spheres are completely smooth so that the inelasticity of
collisions is characterized by a (constant) coefficient of normal
restitution $\alpha \leq 1$. We also assume that the density $n$ is sufficiently low so that the one-particle distribution function $f({\bf r},{\bf v},t)$ of grains verifies the (inelastic) BE. In the absence of external forces, the BE reads
\begin{equation}
\partial_{t}f+\mathbf{v}\cdot \mathbf{\nabla}f=J\left[{\bf v}|f,f\right], \label{1.1}
\end{equation}
where the Boltzmann collision operator $J[f,f]$ is
\begin{equation}
\label{1.2}
J\left[{\bf v}_{1}|f,f\right] =\sigma^{d-1}\int \dd{\bf v}
_{2}\int \dd\widehat{\boldsymbol{\sigma}}\,\Theta (\widehat{{\boldsymbol {\sigma }}}
\cdot {\bf g}_{12})(\widehat{\boldsymbol {\sigma }}\cdot {\bf g}_{12})\left[ \alpha^{-2}
f({\bf r}, {\bf v}_1';t)f({\bf r}, {\bf v}_2';t)-
f({\bf r}, {\bf v}_1;t)f({\bf r}, {\bf v}_2;t)\right].
\end{equation}
Here, $d$ is the dimensionality of the system, $\widehat{\boldsymbol {\sigma}}$ is a unit vector along the line of centers, $\Theta $ is the Heaviside step function, and ${\bf g}_{12}={\bf v}_{1}-{\bf v}_{2}$ is the relative velocity. The primes on the velocities in Eq.\ \eqref{1.1} denote the initial
values $\{{\bf v}_{1}^{\prime}, {\bf v}_{2}^{\prime }\}$ that lead to
$\{{\bf v}_{1},{\bf v}_{2}\}$ following a binary collision:
\begin{equation}
{\bf v}_{1}^{\prime}={\bf v}_{1}-\frac{1}{2}\left( 1+\alpha^{-1}\right)
(\widehat{{\boldsymbol {\sigma }}}\cdot {\bf g}_{12})\widehat{{\boldsymbol {\sigma }}}
,\quad {\bf v}_{2}^{\prime }={\bf v}_{2}+\frac{1}{2}\left( 1+\alpha^{-1}\right)
(\widehat{{\boldsymbol {\sigma }}}\cdot {\bf g}_{12})\widehat{
\boldsymbol {\sigma}}.
\label{3}
\end{equation}

The exact macroscopic balance equations for $n(\mathbf{r},t)$, $\mathbf{U}(\mathbf{r},t)$ and $T(\mathbf{r},t)$ follow directly from the BE \eqref{1.1} by multiplying with $1$, $m\mathbf{v}$, and $\frac{1}{2}mv^2$ and integrating over $\mathbf{v}$. They are given by \cite{BDKS98}
\begin{equation}
\label{4}
D_t n+n\nabla \cdot \mathbf{U}=0,
\end{equation}
\begin{equation}
\label{5}
\rho D_t U_i+\partial_j P_{ij}=0,
\end{equation}
\begin{equation}
\label{6}
D_t T+\frac{2}{dn}\left(\partial_i q_i+ P_{ij}\partial_j U_i \right)=-\zeta T,
\end{equation}
where $D_t\equiv \partial_t+\mathbf{U}\cdot \nabla$ is the material derivative, $\rho=mn$ is the mass density,
\begin{equation}
\label{7}
P_{ij}=\int\; \dd\mathbf{v} \; m V_i V_j f(\mathbf{v})
\end{equation}
is the pressure tensor,
\begin{equation}
\label{8}
\mathbf{q}=\int\; \dd\mathbf{v} \; \frac{m}{2}V^2 \textbf{V}  f(\mathbf{v}),
\end{equation}
is the heat flux and
\begin{equation}
\label{8.1}
\zeta=-\frac{m}{dnT}\int\; \dd\mathbf{v}\; V^2 J[f,f]
\end{equation}
is the cooling rate characterizing the rate of energy dissipated due to collisions. In the above equations, $\mathbf{V}=\mathbf{v}-\mathbf{U}$ is the peculiar velocity.

\section{Grad's moment method}
\label{sec3}

As mentioned in the Introduction, Grad's moment method is based on the expansion of the velocity distribution function in a complete set of orthogonal polynomials (generalized Hermite polynomials), the coefficients being the corresponding velocity moments. In addition, to solve the (infinite) hierarchy of moment equations, the expansion is truncated after a certain order $k$ and so, the above hierarchy becomes a closed set of coupled equations. In the
standard Grad's moment method, the retained moments are the hydrodynamic fields ($n$, $\mathbf{U}$, and $T$) plus the irreversible momentum and heat fluxes ($P_{ij}-p\delta_{ij}$ and $\mathbf{q}$). In the three-dimensional case ($d=3$), there are 13 moments involved in the form of the velocity distribution function $f$; hence this method is referred to as the 13-moment method. In the case of a general dimensionality $d$ the number of moments
is $d(d + 5)/2 + 1$.

On the other hand, since we are interested in comparing the present results with those obtained for granular gases \cite{BDKS98} from the Chapman-Enskog method, I'll include the full contracted moment of fourth order
\begin{equation}
\label{11}
c=\frac{8}{d(d+2)}\left[\frac{m^2}{4nT^2}\int\; \dd\mathbf{v}\; V^4  f(\mathbf{V})-\frac{d(d+2)}{4}\right].
\end{equation}
The inclusion of the scalar field $c$ to the thirteen moments of mass
density, velocity, pressure tensor and heat flux vector will allow us to make a close comparison with the previous forms derived for the cooling rate and the NS transport coefficients \cite{BDKS98}. Thus, the explicit form of the non-equilibrium distribution function for the fourteen moments can be written as
\begin{eqnarray}
\label{9}
f(\mathbf{V})&\to& f_\text{M}(\mathbf{V}) \left\{1 +\frac{m}{2T}\left(P_{ij}-p\delta_{ij}\right)V_iV_j+\frac{2}{d+2}\frac{m}{nT^2}\left(
\frac{m V^2}{2T}-\frac{d+2}{2}\right){\bf V}\cdot {\bf q}\right. \nonumber\\
& & \left. +\frac{c}{4}\left[\left(\frac{mV^2}{2T}\right)^2-\frac{d+2}{2}
\frac{mV^2}{T}+\frac{d(d+2)}{4}\right]\right\},
\end{eqnarray}
where
\begin{equation}
\label{10}
f_\text{M}(\mathbf{V})=n\left(\frac{m}{2\pi T}\right)^{d/2}e^{-mV^2/2T}
\end{equation}
is the local equilibrium distribution function. The coefficients appearing in each one of the velocity polynomials in \eqref{9} have been chosen by requiring that the basic fields, the pressure tensor, the heat flux vector and the fourth moment of the trial function \eqref{9} to be the same as those for the exact velocity distribution function $f$. It must be remarked that the fourth moment $c$ characterizes the deviations of $f$ from its Gaussian form in the homogenous cooling state. Note that, for the sake of simplicity, I have not included all polynomials of fourth order in the trial function \eqref{9}. The inclusion of these moments would modify for instance the form of the cooling rate. However, as mentioned before, I want to offer here a theory with the same degree of accuracy as the one reported \cite{BDKS98} by using the Chapman-Enskog method and so, only non-Gaussian corrections to the homogeneous cooling state distribution will be included in the Grad's distribution function \eqref{9}.

The cooling rate $\zeta$ can be evaluated when one makes use of the form \eqref{9} for $f$ in the definition \eqref{8.1}. Neglecting nonlinear terms in $P_{ij}-p\delta_{ij}$, $\mathbf{q}$, and $c$, the result is
\begin{equation}
\label{12} \zeta=\frac{2}{d}\frac{\pi^{\left( d-1\right) /2}}
{\Gamma \left( \frac{d}{2}\right)}(1-\alpha^2)\left(1+\frac{3}{32}c\right)
n\sigma^{d-1}\sqrt{\frac{T}{m}}.
\end{equation}
The goal now is to determine the first order contributions to the irreversible momentum and heat fluxes and the fourth velocity moment $c$. Let us evaluate each quantity separately.

\subsection{Pressure Tensor}

The pressure tensor is defined by Eq.\ \eqref{7}. In order to evaluate it, I multiply both sides of Eq.\ \eqref{1.1} by $m V_i V_j$ and integrate over velocity. The result is
\begin{equation}
\label{13}
\partial_t P_{ij}+P_{ij} \nabla \cdot \mathbf{U}+U_k \partial_k P_{ij}+P_{kj}\partial_k U_i +
P_{ki}\partial_k U_j +\frac{2}{d+2}\partial_k(q_i\delta_{jk}+q_j\delta_{ik}+q_k\delta_{ij})=
-\nu_\eta (P_{ij}-p\delta_{ij})-\zeta p \delta_{ij},
\end{equation}
where use has been made of the relations
\begin{equation}
\label{16}
\int\; \dd\mathbf{v}\; m V_i V_j V_k f(\mathbf{V})\to \frac{2}{d+2}(q_i\delta_{jk}+q_j\delta_{ik}+q_k\delta_{ij}),
\quad \int\; \dd\mathbf{v}\; m V_i V_j  J[f,f]\to -\nu_\eta (P_{ij}-p\delta_{ij})-\zeta p \delta_{ij},
\end{equation}
where \cite{BDKS98,G02}
\begin{equation}
\label{14}
\nu_\eta=\frac{3}{4d}\left(1-\alpha+\frac{2}{3}d\right)(1+\alpha)
\left(1-\frac{c}{64}\right)\nu_0,
\end{equation}
and
\begin{equation}
\label{15} \nu_0=\frac{8}{d+2}\frac{\pi^{\left( d-1\right) /2}}
{\Gamma \left( \frac{d}{2}\right)}n\sigma^{d-1}\sqrt{\frac{T}{m}}
\end{equation}
is an effective collision frequency related to the NS shear viscosity $\eta_0$  of an elastic gas ($\nu_0=p/\eta_0$). Note that nonlinear terms in $c$, $P_{ij}-p\delta_{ij}$ and $\mathbf{q}$ have been neglected in the collision integral \eqref{16} when $f$ is replaced by its Grad's approximation \eqref{9}.

We are interested in the solution of Eq.\ \eqref{13} in the NS approximation, namely, when the pressure tensor and the heat flux  are given by the constitutive equations  \eqref{1} and \eqref{2}, respectively. Thus, in order to solve Eq.\ \eqref{13}, we need to make use of the balance equations \eqref{4}--\eqref{6} up to first order:
\begin{equation}
\label{18}
\partial_t n \to -\mathbf{U}\cdot \nabla n-n \nabla \cdot \mathbf{U}, \quad
\partial_t \mathbf{U} \to - \mathbf{U}\cdot \nabla \mathbf{U}-\rho^{-1}\nabla p,\quad
\partial_t T \to -\mathbf{U}\cdot \nabla T-\frac{2}{dn}p \nabla \cdot \mathbf{U}-\zeta T.
\end{equation}
The shear viscosity $\eta$ can be easily obtained from Eq.\ \eqref{13} when one takes into account Eqs.\ \eqref{1} and \eqref{18}. The corresponding equation for $\eta$ is
\begin{equation}
\label{19}
\left(\partial_t+\nu_\eta\right) \eta=p,
\end{equation}
where the time derivative $\partial_t \eta$ must be evaluated to zeroth-order in spatial gradients. From dimensional analysis $\eta \propto T^{1/2}$ and so, $\partial_t \eta=\frac{1}{2}\eta \partial_t \ln T=-\frac{1}{2}\zeta \eta$. With this result, the solution to Eq.\ \eqref{19} can be written as
\begin{equation}
\label{22}
\eta=\frac{\eta_0}{\nu_\eta^*-\frac{1}{2}\zeta^*},
\end{equation}
where $\eta_0=p/\nu_0$, $\nu_\eta^*\equiv \nu_\eta/\nu_0$, and $\zeta^*\equiv \zeta/\nu_0$. The expression \eqref{22} for $\eta$ agrees with the one derived \cite{BDKS98} from the Chapman-Enskog method.

\subsection{Heat Flux}

The evaluation of the heat flux follows similar mathematical steps as those made before for the pressure tensor. In order to determine it, one multiplies both sides of Eq.\ \eqref{1.1} by $\frac{m}{2} V^2 V_i$ and integrates over velocity. The result is
\begin{eqnarray}
\label{23}
& & \partial_t q_i+\frac{d}{2}p\partial_t U_i+P_{ij}\partial_t U_j+
\frac{d+2}{4}\frac{c}{m}\delta_{ij}\partial_j(p T)+\partial_j\frac{p}{\rho}\left(\frac{d+4}{2}P_{ij}
-p\delta_{ij}\right)+q_i \nabla \cdot {\bf U}+{\bf U}\cdot \nabla q_i \nonumber\\
& & +\frac{2}{d+2}\partial_j U_k \left(q_i\delta_{jk}+q_j\delta_{ik}+q_k\delta_{ij}\right)
+{\bf U}\cdot \nabla U_j P_{ij}+{\bf q}\cdot \nabla U_i+\frac{d}{2}p{\bf U}\cdot \nabla U_i=
-\nu_\kappa q_i,
\end{eqnarray}
where use has been made of \eqref{16} and the relations
\begin{equation}
\label{24}
\int\; \dd\mathbf{v}\; \frac{m}{2} V^2 V_i V_j  f(\mathbf{V})\to
\frac{d+2}{4}\frac{pT}{m}c\delta_{ij}+\frac{p}{\rho}\left(\frac{d+4}{2}P_{ij}
-p\delta_{ij}\right), \quad \int\; \dd\mathbf{v} \frac{m}{2} V^2 \mathbf{V}   J[f,f]\to -\nu_\kappa \mathbf{q},
\end{equation}
where
\begin{equation}
\label{25}
\nu_\kappa=
\frac{1+\alpha}{d}\left[\frac{d-1}{2}+\frac{3}{16}(d+8)
(1-\alpha)+\frac{4+5d-3(4-d)\alpha}{1024}c\right]\nu_0.
\end{equation}
As before, I have only retained linear terms in $c$, $P_{ij}-p\delta_{ij}$ and $\mathbf{q}$ in the second identity of Eq.\ \eqref{24}. In the NS approximation, Eq.\ \eqref{23} becomes
\begin{equation}
\label{26}
\partial_t q_i-\frac{d+2}{2}\frac{p}{\rho}\partial_i p+\frac{d+2}{2m}\left(1+\frac{c}{2}\right)\partial_i (pT)=-\nu_\kappa q_i.
\end{equation}
Dimensional analysis shows that $\kappa \propto T^{1/2}$ and $\mu \propto T^{3/2}$. Thus, according to the constitutive equation \eqref{2}, the time derivative of the heat flux can be written as
\begin{equation}
\label{27}
\partial_t q_i=2\zeta \kappa \partial_i T+\zeta\left(\frac{T\kappa}{n}+\frac{3}{2}\mu \right)\partial_i n,
\end{equation}
where use has been made of the relation
\begin{equation}
\label{28.1}
\partial_i (\partial_tT)=-\partial_i(\zeta T)=-\zeta T\partial_i \ln n-\frac{3}{2}\zeta T \partial_i \ln T.
\end{equation}
Substitution of Eq.\ \eqref{27} into Eq.\ \eqref{26} allows one to identify the explicit forms of the transport coefficients $\kappa$ and $\mu$. They are given by
\begin{equation}
\label{28}
\kappa=\frac{d-1}{d}\kappa_0\frac{1+c}{\nu_\kappa^*-2\zeta^*},
\end{equation}
\begin{equation}
\label{29}
\mu=\frac{\kappa_0 T}{n}\left(\nu_\kappa^*-\frac{3}{2}\zeta^*\right)^{-1}\left(\kappa^*\zeta^*+
\frac{d-1}{2d}c\right),
\end{equation}
where
\begin{equation}
\label{30}
\kappa_0=\frac{d(d+2)}{2(d-1)}\frac{\eta_0}{m}
\end{equation}
is the low density value of the thermal conductivity of an elastic gas, $\nu_\kappa^*\equiv \nu_\kappa/\nu_0$, and $\kappa^*\equiv \kappa/\kappa_0$. As in the case of the shear viscosity, Eqs.\ \eqref{28} and \eqref{29} for $\kappa$ and $\mu$, respectively, agree with those previously derived \cite{BDKS98} from the Chapman-Enskog method in the first Sonine approximation.

\subsection{Fourth Velocity Moment}

To close the complete determination of the NS transport coefficients, it still remains to evaluate the contracted fourth moment $c$ defined by Eq.\ \eqref{11}. To get it, I multiply now both sides of Eq.\ \eqref{1.1} by $V^4$ and integrates over velocity. In the NS order, one gets
\begin{equation}
\label{31}
d(d+2)\left(1+\frac{c}{2}\right)m^{-2}\left(\partial_t (pT)+\frac{d+4}{d}p T \nabla \cdot \mathbf{U}+
\mathbf{U}\cdot \nabla (p T)\right)=\int\;d\mathbf{v} V^4 J[f,f].
\end{equation}
Equation \eqref{31} can be simplified when one takes into account the balance equations \eqref{18} and the relation (which only applies to linear order in $c$) \cite{NE98}
\begin{equation}
\label{32}
\int\;\dd\mathbf{v}\; V^4 J[f,f] \to -\frac{d+2}{2}\frac{pT}{m^2} (1-\alpha^2)\left[d+\frac{3}{2}+\alpha^2
+\frac{c}{2}\left(\frac{3}{32}(10d+39+10\alpha^2)+\frac{d-1}{1-\alpha}\right)\right]\nu_0
.
\end{equation}
In dimensionless form, Eq.\ \eqref{31} can be written as
\begin{equation}
\label{33}
4d\left(1+\frac{c}{2}\right)\zeta^*=(1-\alpha^2)\left[d+\frac{3}{2}+\alpha^2
+\frac{c}{2}\left(\frac{3}{32}(10d+39+10\alpha^2)+\frac{d-1}{1-\alpha}\right)\right].
\end{equation}
The solution of Eq.\ \eqref{33} can be easily obtained when one uses the expression \eqref{12} for the cooling rate $\zeta$. Retaining only linear terms in $c$, the final form of $c$ is
\begin{equation}
\label{39}
c=\frac{32(1-\alpha)(1-2\alpha^2)}{9+24d-\alpha(41-8d)+30(1-\alpha)\alpha^2}.
\end{equation}
Equation \eqref{39} coincides with the one derived by van Noije and Ernst \cite{NE98} in the free cooling state.

\section{Comparison with previous results}
\label{sec4}

The results derived in the previous Section from Grad's moment method have shown that the explicit forms of the NS transport coefficients are the same as those previously obtained in Ref.\ \cite{BDKS98} by using the Chapman-Enskog method. This shows the equivalence between both methods in the NS regime. On the other hand, as said in the Introduction, Grad's 13-moment method \cite{G49} was already used many years ago by Jenkins and Richman to determine the NS transport coefficients of a dense gas of inelastic hard disks ($d=2$) \cite{JR85a} and spheres ($d=3$) \cite{JR85b}. In the low-density regime and for smooth spheres, their expressions are given by
\begin{equation}
\label{40}
\eta_\text{JR}=\frac{8\eta_0}{(7-3\alpha)(1+\alpha)} \quad (d=2), \quad
\eta_\text{JR}=\frac{4\eta_0}{(3-\alpha)(1+\alpha)} \quad (d=3),
\end{equation}
\begin{equation}
\label{41}
\kappa_\text{JR}=\frac{2\kappa_0}{(1+\alpha)
\left[1+\frac{15}{4}(1-\alpha)\right]} \quad (d=2), \quad
\kappa_\text{JR}=\frac{2\kappa_0}{(1+\alpha)
\left[1+\frac{33}{16}(1-\alpha)\right]} \quad (d=3),\quad
\mu_\text{JR}=0.
\end{equation}
A more recent derivation \cite{K11} of the NS transport coefficients for hard spheres has been also carried out from Grad's fourteen moment theory. It differs from the Jenkins and Richman approach \cite{JR85a,JR85b} by the inclusion of the contracted fourth moment $c$. In particular, the coefficient $\mu$ is different from zero but very small since it is proportional to $c$.
\begin{figure}
\begin{tabular}{lr}
\resizebox{7cm}{!}{\includegraphics{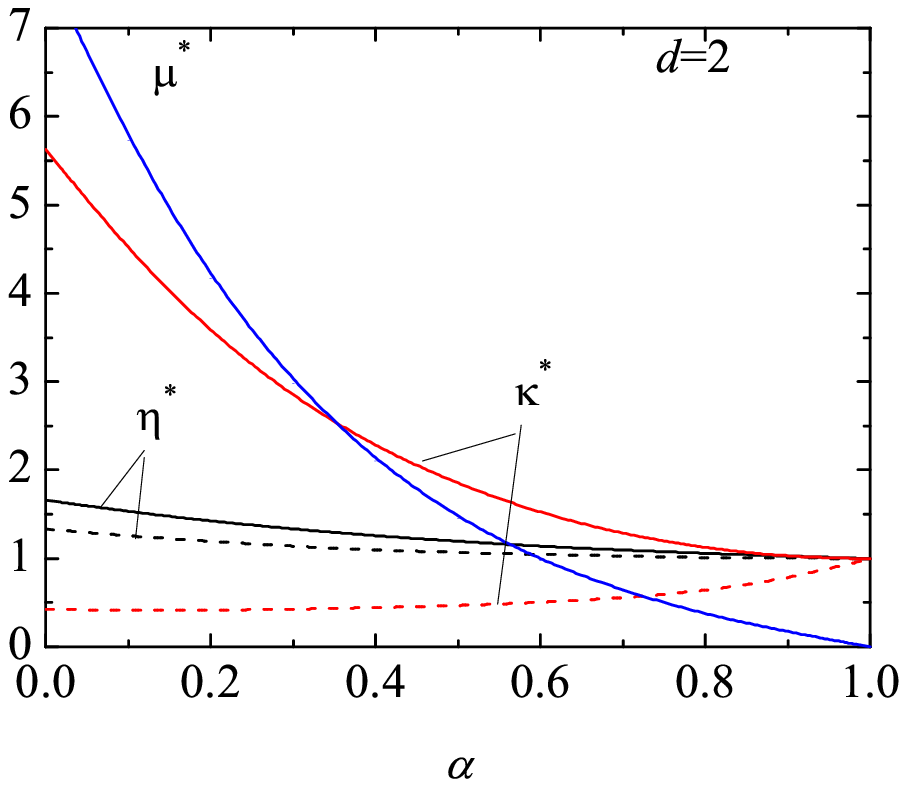}}
&\resizebox{6.9cm}{!}{\includegraphics{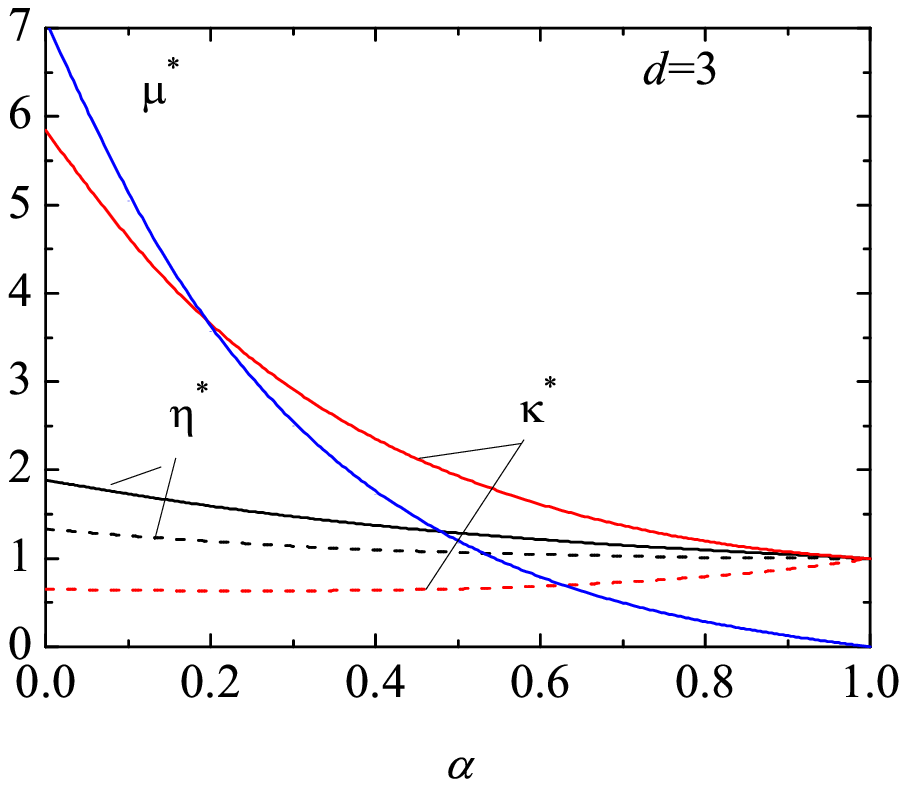}}
\end{tabular}
\caption{Plot of the reduced transport coefficients $\eta^*(\alpha)$, $\kappa^*(\alpha)$,
and $\mu^*(\alpha)$ versus the coefficient of restitution for hard disks ($d=2$)
and hard spheres ($d=3$). The solid lines are the results derived here while the dashed lines
are the results obtained by Jenkins and Richman for disks \cite{JR85a} and spheres \cite{JR85b}. \label{fig1}}
\end{figure}

Comparison between Eqs.\ \eqref{22} and \eqref{40} for $\eta$ and Eqs.\ \eqref{28}--\eqref{29} and \eqref{41} for $\kappa$ and $\mu$ show that the results derived in the present paper differ from those obtained by Jenkins and Richman \cite{JR85a,JR85b}. The differences between both works are due essentially to the assumptions made in Refs.\ \cite{JR85a,JR85b} since the latter authors neglect the time dependence of temperature due to collisional cooling (which is equivalent to take $\zeta^*=0$ in our expressions) and the non-Gaussian corrections to the homogeneous cooling state (which is equivalent to take $c=0$ in our expressions). While the latter simplification is in general not relevant (except for quite extreme small values of $\alpha$), the former hypothesis turns out to be important for finite values of dissipation (i.e., beyond the quasielastic limit $\alpha \simeq 1$). In particular, the coefficient $\mu$ vanishes in the Jenkins-Richman theory \cite{JR85a,JR85b}.

In order to illustrate the quantitative differences between both theories, Fig.\ \ref{fig1} shows the $\alpha$-dependence of the (reduced) transport  coefficients $\eta^*(\alpha)\equiv \eta(\alpha)/\eta_0$, $\kappa^*(\alpha)\equiv \kappa(\alpha)/\kappa_0$, and $\mu^*(\alpha)\equiv n\mu(\alpha)/T\kappa_0$ for hard disks ($d=2$) and spheres ($d=3$). It is quite apparent that while the dependence of the shear viscosity on the coefficient of restitution is relatively well captured by the Jenkins-Richman theory, there are significant differences for the heat flux transport coefficients. In particular, the present theory shows that $\kappa^*$ increases with decreasing $\alpha$ while the opposite happens in the Jenkins-Richman theory. The discrepancies are much more important for the coefficient $\mu$ since my results show clearly that $\mu \neq 0$ in contrast to the prediction of the Jenkins-Richman theory ($\mu_\text{JR}=0$). The comparison with the results derived in Ref.\ \cite{K11} are not presented here since the latter results slightly differ from those obtained in the Jenkins-Richman theory for spheres \cite{JR85b}. It is important to remark that the Chapman-Enskog expressions for $\eta^*$, $\kappa^*$ and $\mu^*$ (which are the same as those obtained here) compare quite well with numerical results \cite{BR04} obtained by solving the BE by means of the direct simulation Monte Carlo (DSMC) method.

In summary, Grad's moment method has been used to determine the NS transport coefficients of a granular gas described by the inelastic BE. As in a previous work \cite{K11}, the present method differs from the conventional 13-moment method by the inclusion of the full contracted fourth moment $c$, defined by Eq.\ \eqref{11}. In contrast to previous attempts \cite{JR85a,JR85b}, the present theory takes into account the time dependence of the granular temperature due to cooling effects as well as the non-Gaussian corrections to the homogeneous cooling reference state. The explicit expressions for the NS transport coefficients derived here agree with those obtained years ago from the Chapman-Enskog expansion in the first Sonine approximation \cite{BDKS98}. This agreement shows the equivalence between both approximate methods to solve the BE for granular gases in the NS regime. The inclusion of more velocity moments (for instance, all the fourth degree velocity moments) would change the final results since there would be likely additional contributions to the cooling rate, for instance. As a future work, I plan to extend the present theory to moderate dense gases \cite{GD99} by using the Enskog kinetic equation.


\acknowledgments
This work has been supported by the Spanish Government through Grant No. FIS2010-16587, partially financed by FEDER funds and by the Junta de Extremadura (Spain) through Grant No. GR10158.


\end{document}